\documentstyle[12pt,aasms4]{article}

\lefthead{Cha, Sembach, \& Danks}
\righthead{Vela Supernova Remnant}

\begin{document}

\newcommand{\degr}{$^{\circ}$}
\newcommand{\kms}{\,km\,s$^{-1}$}     
\newcommand{\vlsr}{V$_{\sc LSR}$}
\newcommand{\updag}{$^\dagger$}

\title{The Distance to the Vela Supernova Remnant}

\author{Alexandra N. Cha\altaffilmark{1}, Kenneth R. Sembach\altaffilmark{1}, Anthony C. Danks\altaffilmark{2}}
\altaffiltext{1}{Department of Physics \& Astronomy, The Johns Hopkins University,
Baltimore, MD  21218}
\altaffiltext{2}{Raytheon STX, Goddard Space Flight Center, Code 683, Greenbelt, Maryland 20771}

\begin{abstract}

We have obtained high resolution \ion{Ca}{2} and \ion{Na}{1} absorption 
line spectra toward 68 OB stars in the direction of the Vela Supernova
Remnant. The stars lie at distances of 190 -- 2800 pc as determined by 
Hipparcos and spectroscopic parallax estimations. The presence 
of high velocity absorption attributable to the remnant along some of the 
sight lines constrains the 
remnant distance to 250$\pm$30 pc. This distance is consistent with several 
recent investigations that suggest that the canonical remnant 
distance of 500 pc is too large.
\end{abstract}

\keywords{line: profiles -- ISM: clouds --
ISM: individual (Vela Supernova Remnant) --
ISM: supernova remnants}

\section {Introduction}

Supernova remnant (SNR) studies provide valuable information about
the amount of energy released in supernova explosions, the effects of 
blast waves on the structure and physical conditions in the interstellar
medium (ISM), shock physics, and dust grain destruction.  Distance is a 
fundamental quantity that often factors into such studies.  
Knowledge of SNR distances is crucial 
for determining and justifying calculations of remnant ages 
(Wallerstein \& Silk 1971), predicting
the explosion energies of the progenitor supernovae, and quantifying 
temperatures behind shock fronts (Jenkins \& Wallerstein 1995). 

The Vela SNR is one of the best studied SNRs in the sky.  An uncertain 
distance of 500 pc (Milne 1968) is often quoted.  Recent studies
of the conditions within the remnant suggest that a smaller
value is more appropriate.  We have undertaken an observing program to 
determine the distance to the Vela SNR directly by spectroscopic means.  
We present these results in this {\it Letter} and provide additional 
results pertaining to time variability in the high velocity
absorption in a companion paper (Cha \& Sembach 1999, hereafter Paper II).

\section{Data}
We obtained high-resolution \ion{Ca}{2} K (3933.663 \AA) and 
\ion{Na}{1} D$_2$,D$_1$ (5889.951, 5895.924 \AA)
spectra of 68 OB stars in the direction of the Vela SNR
using the Coud\'e Echelle Spectrograph 
on the Coud\'e Auxiliary Telescope at the European Southern Observatory in 
1993, 1994, and 1996.  The instrumental setups
are described in Paper II along with the data reduction procedures (see also
 Sembach et al. 1993).
The fully reduced data have typical signal-to-noise ratios in 
excess of 100 and 2-pixel resolutions of 
$\lambda/\Delta\lambda$\,$\approx$\,75,000, or $\approx$4 \kms, 
as determined from the widths of Th-Ar 
lamp spectra.  The spectra are presented in Paper II.

The stars observed rotate rapidly and serve as suitable
background sources for studies of narrow interstellar (or remnant) absorption
lines.  The locations of the stars are shown in Figure~1 along with
a ROSAT 0.75 keV X-ray contour 
(Snowden et al. 1995) arising from the extreme edge of the adiabatically
expanding shock encircling the remnant.    

\section{Previous Distance Estimates}
The distance to the Vela SNR has been questioned ever since 
Milne (1968) proposed a distance of 500$\pm$100 pc.  Milne arrived at this 
value by comparing the observed angular diameters of the Vela SNR,
Cygnus Loop, and IC~443, and assuming that all three had similar 
linear diameters.  Comparing the Cygnus Loop ($\phi$ = 170\arcmin, 
d\,=\,770 pc; Minkowski 1958; Harris 1962) to Vela, ($\phi$ = 4\degr), 
yielded d$_{Vela}$\,$\approx$\,540 pc.  A similar comparison of Vela to 
IC\,443
($\phi$ = 45\arcmin, d\,= 2 kpc -- Milne's ``probable distance'') resulted in
d$_{Vela}$\,$\approx$\,375 pc.  The Vela SNR distance of 
500$\pm$100 pc thus arose from a rough average of these distances.
Kristian (1970) summarizes five early distance measurements,
similarly concluding that d$_{Vela}$=500 pc, but with a factor of two
uncertainty. 

A compilation of recent distance estimates to the Vela SNR
can be found in Table~1.  
The methods 
included a repeat of Milne's calculations 
using an updated angular diameter of the
Vela SNR (Oberlack et al. 1994).  That work, which 
also nicely recounts previous distance determinations, and suggests
d$_{Vela}$\,=\,230$\pm$$_{115}^{230}$ pc. 
A second distance estimate involved an analysis of the proper motion
of the Vela pulsar by {\"O}gelman et al. (1989), who found that if the
distance to the pulsar were d$_{Vela}$\,= 290$\pm$80 pc,
then the projected velocity of the pulsar matched the
pulsar velocity derived from interstellar scintillation data (Cordes 1986).  

As early as 1971, Wallerstein \& Silk (1971) recommended a distance 
near 250 pc to unify age estimates of the remnant 
derived from the observed rate of pulsar deceleration 
with estimates based on the shock velocity and angular size of the remnant.
More recently, Jenkins \& Wallerstein (1995)
quantified the total energy of the Vela SNR and found that 
a distance of 500 pc resulted in an energy that was an order
of magnitude larger than the expected $\approx$\,10$^{51}$ergs
for Type II supernovae (Woosley \& Weaver 1986; Arnett et al. 1989).  
A factor of 2 smaller distance lowered the energy by a factor of 8
to a more representative value.
Jenkins \& Wallerstein (1995) also noted  
the high correlation between the theoretically 
calculated temperature behind the shock wave and the 
observational temperature estimates of the X-ray
emitting regions of Vela (Kahn et al. 1985) if the adopted distance to Vela 
were $\approx$\,250 pc.   

\section{A Precise Distance to the Vela SNR}

We have analyzed high-resolution \ion{Ca}{2} absorption spectra of 
68 stars in the direction of the Vela SNR
($l$\,$\approx$\,264\degr, $b$\,$\approx$\,$-$3\degr).
Many of the spectra reveal 
\ion{Ca}{2} and \ion{Na}{1} absorption at velocities well beyond those 
expected for gas in the local interstellar medium (LISM), which
has \vlsr(\ion{Ca}{2})$\approx$\,0.9$\pm$9.4 \kms
(Vallerga et al. 1993).   
In the direction
of Vela, we find no ISM
absorption toward $\delta$~Vel (d\,$\approx$\,20 pc).
Crawford (1991) found interstellar \ion{Ca}{2} at 
velocities of --6.6 and --1.6 \kms\ 
toward $\kappa$~Vel (d\,$\approx$\,100 pc).  
These velocities are typical of many southern hemisphere (Sco-Cen-Vel)
sight lines extending 100--200 pc from the Sun; the southern 
hemisphere LISM components span the --20 to +6 \kms\ velocity range 
(Crawford 1991; Welsh et al. 1994; G\'enova et al. 1997).
In contrast to these
relatively low LISM mean velocities, Dubner et al. (1998) reported the
presence of \ion{H}{1} 21cm emission from gas about the center of the SNR
accelerated to velocities
of $|$\vlsr$|$ $\approx$\,30 \kms\ by the expansion of the Vela 
supernova shock.  

To determine the distance to the Vela SNR, we decomposed each 
\ion{Ca}{2} profile into Gaussian components and identified 
those components that appear at velocities indicative of the remnant. 
In Figure~2 we plot the central velocities of each component as a 
function of background star distance.  To illustrate the characteristically
low velocity foreground gas at 100 pc $<$ d $<$ 200 pc, we also plot
data for stars in the Sco-Cen region (Crawford 1991).
We estimated stellar distances and errors using two methods: 
trigonometric parallaxes based upon Hipparcos parallax 
measurements and 
spectroscopic 
parallaxes based upon photometric colors and spectral types reported in the 
literature
(see Paper II).  There was good agreement between the two methods, but
whenever possible, we
adopted the more accurate Hipparcos trigonometric parallax values.

Given the systematically low velocities of LISM material and the 
higher velocity \ion{H}{1} emission seen by Dubner et al. (1998) toward
Vela, we adopted a lower limit of $\pm$25 \kms\ for absorption that we 
considered to be associated with the remnant.  A much more conservative limit
of $\pm$50 \kms\ could also be adopted, and for the sake of erring on the side
of caution we will discuss our distance estimate with this higher velocity
cutoff when appropriate.   These velocities are highlighted by dashed and solid
vertical lines, respectively, in Figure~2.  Beyond 250 pc, all of our
\ion{Ca}{2} spectra but one (HD\,74662) show absorption components at
$|$\vlsr$|$$>$ 25 \kms,
and beyond 390 pc absorption lines with $|$\vlsr$|$ $>$ 50 \kms\ 
appear in many of the spectra.

We show the \ion{Ca}{2} K spectrum of two stars at 250 pc, HD\,70309
(d$_{Hipp}$ = 250$\pm$35 pc) and HD\,71459
(d$_{Hipp}$ = 250$\pm$30 pc), in Figure~3.  
Three absorption components at 
$<$\vlsr$>$ = $-$28, $-$13, and +1 \kms\ are readily discernible
in the spectrum of HD\,70309.
The $-$13 and +1 \kms\ components are also present in the spectrum of
nearer stars and
probably arise in the LISM at a distance of 40--200 pc.  We believe 
the $-$28 \kms\ component
reveals gas associated with the front side of the remnant.
The spectrum of HD\,71459 also reveals higher velocity \ion{Ca}{2}
absorption.  Figure 3 shows that there is 
an absorption line with $<$\vlsr$>$ = +30 \kms, indicating 
the presence of remnant gas moving 
away from us. The two sight lines are separated by 6.2\degr, or 
about 27 pc at a distance of 250 pc.  This is roughly the diameter of the 
remnant at this distance.
Both sight lines pass near
the edge of the X-ray emission line boundary of the remnant, so the total
space velocities of the remnant gas along the sight lines could
be considerably higher than the measured velocities.  

We now consider a simple model to determine what the deprojected 
space velocities for these two sight lines might be.
We assume that the remnant is a spherically expanding shell centered
in the X-ray contour, with approximate coordinates 
$\alpha_{2000}$ = 08$^h$\,40$^m$ and
$\delta_{2000}$ = $-$45\degr\,00\arcmin.\footnotemark
 Applying a spherical coordinate system allows the
total space velocity of the gas to be resolved into vector components, 
including the
observed velocity, which is the projection of the total
velocity onto the line of sight.  We assign a right handed
coordinate system with the origin at the center of the remnant and the
following axes: the x-axis extends from the observer to the remnant,
the y-axis is right ascension, 
and the z-axis is declination.  Converting from spherical to
Cartesian coordinates, we find the space velocity, v, knowing the observed
velocity (x-projection), v$_{obs}$,  to be v$_{obs}$ = vsin$\theta$cos$\phi$.
From Figure~1, we estimate $\theta$ $\approx$ 45$^\circ$ for HD\,71459 and 
$\theta$ $\approx$ 130$^\circ$ for HD\,70309.  Determining $\phi$ would
require knowing the distance to clouds on the spherical remnant.  
A robust lower limit on the total velocity is obtained by setting 
$\phi$ = 0$^\circ$ (i.e., v $>$ v$_{obs}$csc$\theta$).

\footnotetext{We note that the \ion{Ca}{2} data is not described
well by a simplistic spherical model (see Wallerstein et al. 1980), so we
use it only as a first order approximation to illustrate the general
effects of projection.}

Applying the projection effects to the absorption components of the two 
stars at 250 pc, we find that the observed velocity of the component at
 $-$28 \kms\ toward HD\,70309 converts to a minimum space velocity
of $-$40 \kms, and the observed velocity of the component at 
+30 \kms\ toward HD\,71459 corresponds to a minimum space velocity
of +39 \kms.  These space velocities are
indeed much higher than those expected for gas associated with the LISM, 
lending
support to our claim that observed LSR velocities  beyond $\pm$25 \kms\
toward these stars arise within the remnant.

Higher velocity absorption is also seen along some sight lines toward
Vela.  Figure~3 contains the \ion{Ca}{2} spectrum of HD\,72350, a B4 IV
star at 
d = 390$\pm$100 pc, as determined using the standard reddening relation,
A$_v$~=3.1~E(B-V), with the UBV photometry of Schild et al. (1983).  
Here, absorption components are observed at 
intermediate velocities, \vlsr\ = $-$20  and +38 \kms, 
as well as at high velocities, \vlsr\ = +114  and +122 \kms. 
The spectrum of another star at a similar distance,
HD\,74319 (d$_{Hipp}$ = 400$\pm$90 pc), also reveals high velocity absorption
components, though at negative velocities, \vlsr\ = $-$61  and $-$74 \kms.  
Both stars are located near the center of
the X-ray shell (Figure~1). The observed absorption velocities along these 
sight lines are probably close to the actual space velocities of the shocked 
regions they probe. 
  
The distribution of sight lines with widely varying absorption signatures
is testimony to the patchiness of the remnant region.
The observed distant and nearby stars are randomly
projected onto the remnant, revealing no obvious bias in their spatial
distribution.
We calculated the two-sided Kolmogorov-Smirnov (K-S) statistic
on the
spectral components velocities for the stars at 245 pc $<$ d $<$ 385 pc and
385 pc $<$ d $<$ 555 pc and found it to be, D=0.14, with a significance
level of 0.40, implying that the two groups of stars are not from significantly
different populations.

Assigning gas with observed velocities of $|$\vlsr$|$ $>$ 25 \kms\
to the Vela SNR results in an SNR distance of 250$\pm$30 pc.  
An upper limit of $\sim$390 pc is allowed if only 
$|$V$_{LSR}|$ $>$ 50 \kms 
components are retained in the analysis. In light of the LISM data, we believe
that the upper limit is overly pessimistic, and therefore we
recommend a distance of 250$\pm$30 pc be used for future studies of the 
remnant.

We report and recalculate some general characteristics of the Vela SNR in
Table~2 using a distance of 250$\pm$30 pc.  The recalculations of the various 
attributes of the remnant follow the prescriptions outlined in the 
previously published work described in \S3.

\section{Supporting Absorption Line Data}
We have reviewed published ultraviolet
absorption line data to test our observationally
determined distance to the Vela SNR.  
Choosing components with $|$\vlsr$|$\,$>$\,25 \kms\
in the International Ultraviolet Explorer (IUE) data
reported by Jenkins et al. (1984)
yields an upper limit of $\sim$400 pc to the distance, based upon an 
uncertain detection of 
\ion{C}{4} at $-$33 \kms\ toward HD\,72555 (d = 390$\pm$90 pc), and the
detection of a 
$-$125 \kms\ absorption component in low ionization lines
(\ion{Fe}{2}, \ion{Si}{2}) toward HD\,74436 (d = 410$\pm$230 pc).  
Copernicus data for stars in the 
direction of the Vela SNR (Wallerstein et al. 1980) 
yield an upper limit of 480$\pm$120 pc as constrained by the 
sight line to HD\, 72127A.  The
distance obtained using the Copernicus data set is almost a factor of two
higher than our measurement, but this may be attributed to the limited sample
of 14 stars, which results in a much coarser sampling of distances.  
The IUE sample was more extensive (45 sight lines) and yields an upper
bound consistent with our distance estimate.  The major
advantage of our spectroscopic investigation
over previous absorption line studies of
the region is the combination of high signal to
noise and high spectral resolution, resulting in a high sensitivity
to weak lines that may not have been observable in the Copernicus and IUE
data.
 
\section{Conclusions}
The analysis of a significant sample of O and B stars 
in the direction of the Vela SNR has enabled a direct distance 
measurement of 250$\pm$30 pc, with a conservative 
upper limit of 390$\pm$100 pc.  This measurement is 
consistent with recent arguments that favor a smaller distance to Vela.

\acknowledgments  
We appreciate comments on this work from our colleagues at the Johns 
Hopkins University.
KRS and ANC acknowledge support from NASA Long Term Space Astrophysics Grant 
NAG5-3485 and grant GO-06413-95A from the Space Telescope Science Institute,
which is operated by AURA under NASA contract NAS5-26555.

\newpage

\begin{center}Figure Captions\end{center}

\noindent
\figcaption{Positions of the sample stars in the sky. The stars
discussed in \S4 are identified. An ``X'' marks the location of the Vela 
pulsar. The ROSAT X-ray contour represents 4 $\times$ 10$^{-4}$ 
counts~s$^{-1}$ arcmin$^{-2}$.}
  
\noindent
\figcaption{Central velocities (diamonds) of the observed
\ion{Ca}{2} absorption 
components embedded with horizontal ticks revealing the Doppler spread
parameters versus the distance of the corresponding star.
Crosses indicate \ion{Ca}{2} velocity data by
Crawford (1991).}

\noindent
\figcaption{\ion{Ca}{2} K absorption spectra towards
HD\,70309, HD\,71459, HD\,72350, and HD\,74319.
Tick marks identify absorption by gas within the SNR.}

\end{document}